\documentclass[
]{ceurart}

\usepackage{hyperref}
\usepackage{csquotes}
\usepackage{graphicx}
\newcommand{\qte}[1]{\textit{``#1''}}

\begin{document}

\copyrightyear{2021}
\copyrightclause{Copyright for this paper by its authors.
  Use permitted under Creative Commons License Attribution 4.0
  International (CC BY 4.0).}

\conference{IntRS'21: Joint Workshop on Interfaces and Human Decision Making for Recommender Systems, September 25, 2021, Virtual Event}

\title{How does the User's Knowledge of the Recommender Influence their Behavior?}


\author[1]{Muheeb Faizan Ghori}[%
email=mghori2@depaul.edu,
]
\address[1]{School of Computing,
  DePaul University, Chicago, USA.}

\author[1]{Arman Dehpanah}[%
email=adehpana@depaul.edu,
]
%\address[1]{School of Computing, DePaul Universtiy, Chicago, USA.}

\author[1]{Jonathan Gemmell}[%
email=jgemmell@cdm.depaul.edu,
]
%\address[1]{School of Computing,
% DePaul University, Chicago, USA.}

\author[1,2]{Hamed Qahri-Saremi}[%
email=hamed.qahri{\char`_}saremi@colostate.edu,
]
\address[2]{College of Business,
 Colorado State University, Colorado, USA.}

\author[1]{Bamshad Mobasher}[%
email=mobasher@depaul.edu,
]
%\address[1]{School of Computing,
 %DePaul University, Chicago, USA.}

\begin{abstract}
Recommender systems have become a ubiquitous part of modern web applications.
They help users discover new and relevant items.
Today's users, through years of interaction with these systems have developed an inherent understanding of how recommender systems function, what their objectives are, and how the user might manipulate them.
We describe this understanding as the \textit{Theory of the Recommender}.
In this study, we conducted semi-structured interviews with forty recommender system users to empirically explore the relevant factors influencing user behavior.
% investigate the various factors influencing user behavior such as their attitude, knowledge, and motivation?
Our findings, based on a rigorous thematic analysis of the collected data, suggest that users possess an intuitive and sophisticated understanding of the recommender system's behavior.
We also found that users, based upon their understanding, attitude, and intentions change their interactions to evoke desired recommender behavior. 
% correct this statement ~~~
Finally, we discuss the potential implications of such user behavior on recommendation performance.
\end{abstract}

\maketitle
\section{Introduction}
\label{sec:Intoduction}
\vspace{-1mm}
% ~ Add a few sentences stating that we explored user attitudes, their perception of how and why the system recommended them items and their behavior?

% Para 1: General Intro
Recommender systems suggest relevant items to users in a variety of domains such as online retailers, streaming services, and social media platforms.
These systems have become an essential tool in modern web applications helping users navigate large and complex online environments.
In domains like e-commerce, recommender systems help service providers boost their revenue and provide a competitive edge. 
These systems often leverage user information and interactions with the system to provide personalized recommendations that satisfy the needs and interests of the user. 

% Para 2: Introducing the context
Recommender systems have become pervasive in the last decade. 
Consequently, users find themselves interacting with recommenders on a regular basis.
The system suggests relevant items to users that satisfy their needs and preferences.
Users view these recommended items, consume items that catch their interests, and perhaps rate or leave feedback about these items.
The system in turn models their responses and provides new recommendations.
However, these repeated interactions may contribute to the user developing an inherent understanding of how recommender systems function.
\looseness=-1

%Para 3: Hypothesis for this research!
Prior research demonstrates that users possess a cognitive model of how recommender systems work~\citep{kulesza2012tell, ngo2020exploring, ghori2019does}.
This cognitive model represents the users' interpretation of how the system operates, what their objectives are, and how they generate recommendations.
For example, users may recognize that the system uses their demographic data and online interactions to build recommendations.
Similarly, they may also recognize that their preferences and interactions influence the recommendations they receive later.
Users may also realize the motives of the system such as increasing revenue, marketing, and advertisements.
However, we hypothesize that such a sophisticated understanding of the system may influence how users choose to interact with it thereby modifying their interactions to obtain desired recommender behavior. 

% Para 4: Why is it important?...
A sophisticated understanding of the recommender system can have several implications.
Kulesza et al.~\citep{kulesza2012tell} showed that a user’s mental model of the recommender system can be helpful in debugging and personalizing the intelligent agent.
In this case, a user unhappy with the current set of recommendations may purposely search for and upvote items they have previously enjoyed in order to improve their user profile and correct the system's reasoning.
This allows the system to revise user preferences and improve recommendations.
On the contrary, users may change their interactions based on certain situations or personal motivations thereby manipulating the feedback that the recommender system receives. 
For example, users, when presented with a political viewpoint with which they disagree, may aggressively downvote a content creator in a video sharing service in order to signal the system that they are uninterested in those viewpoints.
Some users may delete their recent activity on a streaming service to avoid receiving related recommendations.
Other users, sensitive to privacy concerns, may forgo the benefits of a recommender system and opt to view news stories in ‘incognito’ mode.
However, such user behavior may result in an incorrect interpretation of user preferences by the system and have an adverse effect on recommendation performance. 

% Para 5: What do we do in this paper and why is it important?
In this research, we explore how users interact with recommenders based upon their cognitive model of the system. 
We hypothesize that a sophisticated understanding of the recommender system may influence how users choose to interact with it thereby altering the feedback the system receives. 
Understanding such user behavior and the factors affecting it is crucial for several reasons, among them:
1) identifying user behavior and its impact on recommendation performance.
2) designing systems that can leverage these behaviors to improve recommendation performance and the user's satisfaction with the system.
% unintended effect on future recommendations. 
% ~~~~ need to add points here~~~~~

% Para 6: How are we doing it.
To test our hypothesis, we conduct a user study to elicit the users' perception and understanding of how the system operates, their attitude, and their interactions with the system along with the motivation behind them.
To that end, we asked participants to describe recommender systems, how they work, and provide reasoning as to why they may have received certain recommendations. % ~~~~
We then investigate how different users respond to the recommendations to identify unconventional user behavior.
Our results show that users possess a mature understanding of the system's functionality allowing them to reason and predict the recommender's behavior.
We also found that users, based on their personal motivations, modify their interactions to steer the recommender system in desired directions.
The insights obtained from this study provides a refined understanding of how users interact with recommender systems and the various factors that influence user behavior. 

% Finally,we discuss the potential implications of such user behavior on recommendation performance

% The rest of this paper is organized as follows.
% In Section~\ref{sec:RelatedWork} we present our related work.
% A brief summary of common recommendation algorithms and previous work is provided.
% In Section~\ref{sec:Methodology} our methodology is described in detail.
% The results of the study are described in Section~\ref{sec:Results}.
% In section~\ref{sec:Discussion}, we discuss the results in detail.
% Finally, in Section~\ref{sec:Conclusion} we conclude the paper and mention future works.

%---------- ---------- ---------- ---------- ---------- ---------- ---------- ----------
\section{Related Work}
\label{sec:RelatedWork}

Prior research on user perception of the recommender system has shown that users possess a cognitive model of how the system works~\citep{ghori2019does, ngo2020exploring}.
% *(edit)* We speculate that users, based on their understanding of the system, modify their interactions to obtain desired recommender system behavior.
Cognitive models or mental models can be described as the internal representations that users build based on their interactions with the target system~\citep{norman1983some, rumelhart1983representation}.
These models reflect the user's knowledge and beliefs about the system, allowing them to understand, reason, and predict their behavior.  
Mental models vary in their fidelity. 
An ideal model of a target system must be accurate, consistent, and complete. 
However, Norman~\citep{norman1983some} observed that users' mental models of technological systems can be inaccurate, contain areas of uncertainty, or lack completeness. 
Thus, in order to successfully predict a system's behavior, the user's mental model must have some degree of technical validity.
Despite the limitations of mental models, Doyle and Ford~\citep{doyle1998mental} concluded that they are enduring, accessible, and have a great potential to account for human behavioral patterns and errors.
\looseness=-1

Research examining user's mental models in the context of the recommender system is sparse. 
% previous research 1
Kulesza et al.~\citep{kulesza2012tell} conducted an empirical study to explore the effects of users' mental model soundness on personalization in a music recommender system. 
They showed that users with a sound mental model of the recommender system's reasoning can help debug and personalize the intelligent agent.
% previous research 2
In another study, Kulesza et al.~\citep{kulesza2013too} explored ways to improve 
the users' mental model of the system and trust formation using explanations.
They showed that making a recommender system's reasoning transparent using explanations helped users understand the system's reasoning and build trust in the system.
% Previous research 3
In an exploratory study, Ngo et al. ~\citep{ngo2020exploring} elicited mental models of Netflix users and investigated their perception of transparency and control.
The authors propose aligning the explanatory and interactive components of the system with underlying recommendation algorithms and linking the system components to identified basic mental models to increase transparency and control.
% Previous research 4
Eiband et al.~\citep{eiband2018bringing} proposed a stage-based participatory approach for designing transparent recommender system interfaces.
The study provides key insights for practitioners to integrate transparency into recommender system design; achieved by improving the users' mental model of the recommender system through the use of explanation interfaces. 
% Previous research 5
Rader and Gray~\citep{rader2015understanding} investigated user understanding of algorithmic curation in Facebook's news feed.
Analyzing survey responses of Facebook users, they found that users' beliefs about the system varied, and the survey participants demonstrated intuitive theories about how Facebook's news feed works.

As with research on understanding mental models of recommender systems, these studies have not investigated the effect of user's knowledge on their behavior.
On the other hand, research examining mental models in other domains has shown that the user’s knowledge of a target system has the ability to influence user behavior~\citep{zhang2008influence, kang2015my}.
% *(edit)* In a user study, Zhang~\citep{zhang2008influence} explored the effects of mental models on users' web searching behavior. 
% They showed that the familiarity of the task to subjects had a major effect on their web searching behaviors.
% *(edit)* Similarly, Kang et al.~\citep{kang2015my} conducted a qualitative study to understand how users' knowledge of the internet influenced their responses to privacy and security risks.
In this research, we explore the users' perception and beliefs about how recommender systems work in a qualitative study.
Next, we investigate the impact of users' mental models on their interactions with the system. 
Finally, we draw on the `Theory of planned behavior' to understand the key determinants motivating users to such user behavior.

% Linking research to Theory of planned behavior
% *(edit)* To understand user behavior of the recommender system, we draw on the `Theory of Planned Behavior', a theory that provides parsimonious accounts of the determinants of behavior.
The theory of planned behavior~\citep{ajzen1991theory, ajzen1985intentions} states that the performance of a behavior is jointly driven by intentions and perceived behavioral control.
Intentions capture the motivational factors that influence behaviors; factors such as the amount of effort and the degree to which an individual is willing to perform that behavior.
Perceived behavioral control refers to the beliefs regarding the possession of required resources and opportunities to engage in a behavior. 
Most notably, the perception of behavioral control plays an important role, impacting both the intentions and the actual behavior. 
In general, the theory of planned behavior details the determinants of an individual's decision to enact a particular behavior~\citep{conner1998extending}.
This theory has been widely used to understand and predict user behavior in a variety of domains, and to serve as a framework for behavior change interventions~\citep{godin1996theory, conner1998extending}.
Therefore, in this research, we examine how an individual's beliefs about the system's functionality, their attitudes, and intentions affect online behavior. % change to examine relationships?
\looseness=-1

Most popular methods to elicit mental models include diagramming exercises, think-aloud tasks, and verbal interview. 
Carley and Palmquist~\citep{carley1992extracting} proposed that a representation of a mental model can be extracted from verbal text elicited through interviews. 
The identified verbal structure is a symbolic representation of an individual's cognition~\citep{carley1992extracting, fauconnier1994mental, sowa1984conceptual}.
In line with this, we interviewed a sample of recommender system users to elicit their mental models using thematic analysis.
Our work aims to examine the relationship between user’s knowledge and their  behavior, what impact such behavior might have on these systems, and how recommender systems might be designed to cope with, or even leverage these behaviors.

%---------- ---------- ---------- ---------- ---------- ---------- ---------- ----------
\section{Methodology}
\label{sec:Methodology}

The purpose of this study is to identify and investigate the various factors influencing user behavior of the recommender system.
To that end, we used thematic analysis, a well-known qualitative research method.
We conducted in-depth interviews with participants to probe their understanding of the recommender system along with the rationale behind their online interactions.
\looseness=-1
%We chose a qualitative approach for this study since user perceptions can be unique and highly subjective.

% Thematic analysis
\subsection{Thematic Analysis}
Thematic analysis is a well-known qualitative data analysis method for systematically identifying, analyzing, and interpreting themes within qualitative data~\citep{boyatzis1998transforming}.
The identified themes serve as a framework for organizing and reporting meaningful analytic observations relevant to the research question.
Thematic analysis was well suited for our research, as specified by Braun and Clarke~\citep{braun2006using}, this method is especially useful for examining the perspectives of different research participants, highlighting similarities and differences, and generating unanticipated insights.
Thematic analysis has been successfully employed in healthcare and information system domains to uncover user perceptions of health applications and critical user experiences~\citep{kari2016critical}. %with self-tracking technology~\citep{kari2016critical}.

%The resulting finding can be presented using a thematic map.
% ~~~~~~~~~~~~~~~~~ to develop a thematic map based on literature and qualitative results. A thematic map is a graphical tool for “organizing and representing knowledge […and included] concepts, usually enclosed in circles or boxes of some type, and relationships between concepts” (Novak and Cañas 2006, p. 1).
% ~~~~~~~~~~~~~~~~~About thematic map~~~~~~~~~~~~~~~~~~~~~

% Participants
\subsection{Participants}
A total of 40 participants (24 female, 16 male) were recruited for the study using Reddit,  a community-based discussion website.
The age of the participants ranged from 19 to 45 years.
Typically, interviews took between 30-45 minutes and each respondent received a compensation of USD \$10 for their participation.
The participants came from a wide range of professions including a business owner, teacher, sales representative, engineer, and business analyst. 
Participants held diverse educational backgrounds and stated using recommenders for 4.79 hours a day on an average.
%Participants held diverse educational backgrounds ranging from a high school degree to a master’s degree in various disciplines and stated using recommenders for 4.79 hours a day on an average.
The study focused mainly on applications that users frequently use on a daily basis (e.g., e-commerce and streaming platforms such as Amazon, YouTube, Netflix, and Spotify).
\looseness=-1

% Interview Questions: 
\subsection{Study Procedure}
In order to collect rich and detailed qualitative data, we conducted semi-structured interviews using Zoom, a video conferencing service~\citep{archibald2019using}.
Before each interview began, participants were debriefed about the research and their informed consent was obtained as prescribed by the Institutional Review Board.
% *(edit)* We then asked basic demographic information such as age, gender, education, and profession.
The initial interview questions were designed to elicit the user's understanding of the recommender system and investigate their behavior.
Consequently, we probed the participant's perception of how recommenders work. 
Questions included, what is a recommender system, how does the system build recommendations? What information does it use? What are the goals and motivations of the system?  
Similarly, we asked participants to explain the system's rationale behind the various recommendations they receive on their personalized feed.
Finally, the participants were prompted to describe their daily interactions with the system in detail.
Participants answered these questions based on their own experience and understanding, citing examples from the recommenders they use regularly. 
By adapting the conceptual framework of the `theory of planned behavior' to our research context coupled with the insights obtained through initial exploratory interviews with participants, we identified the key determinants affecting user behavior.
The identified factors include the user's overall knowledge of recommenders, behavioral intentions, and attitudes towards the system.
Hence, we explored and investigated the relationships among the user's attitude towards the system, their perceptions and beliefs about how recommenders work, their online behavior, and the motivation behind them.
\looseness=-1
% The interview script was evaluated by a panel of three domain experts in recommender systems to establish content validity and face validity~\citep{boudreau2001validation, fink1995measure}. 

\subsection{Data Analysis}
% Deductive approach - basis TBP
We followed the six-phase approach to thematic analysis as specified by Braun and Clarke~\citep{braun2012thematic, braun2006using}.
Each interview was transcribed and analyzed using QSR NVivo 12~\cite{johnston2006software}.
%, a qualitative data analysis software.
%Each interview was transcribed using a web-based transcription service and analyzed using QSR NVivo 12, a qualitative data analysis software.
First, we performed multiple readings of the textual interview data to gain a comprehensive understanding of the content. 
In the second phase, we systematically analyzed the data by reading the participant responses analytically, and critically to identify initial codes. 
Codes are semantic labels that represent the participant’s interpretation of the data.
Afterward, the identified codes were reviewed to find similarities and differences.
We grouped similar codes into categories known as themes~\citep{braun2012thematic}.
%A theme captures important insights and represents meaningful patterns in the data relevant to the research~\citep{braun2012thematic}.
In the fourth phase, the discovered themes were further condensed into higher-level themes to ensure a distinct and coherent set.
We recursively reviewed and revised the identified themes against the entire data set to verify they adequately capture the data.
In the next phase, we performed a deeper analysis to identify global themes.
We further defined the themes in terms of its focus, scope, and purpose; that each, in turn, builds on and develops the sub-themes.
The resulting global themes were inspired by the conceptual framework of the theory of planned behavior.
This recursive process of thematic analysis resulted in several distinct and coherent themes.
In the last phase, we defined and named the different themes and developed a thematic map that describes the various factors influencing user behavior of the recommender systems.
\looseness=-1

% We present these themes in the results section.
% themes inspired by the theory of planned behavior.

% Using the conceptual framework of the theory of planned behavior as its basis, we 

%---------- ---------- ---------- ---------- ---------- ---------- ---------- ----------
\section{Results}
\label{sec:Results}

\begin{figure}[h]
    \centering
    \includegraphics[scale=0.45]{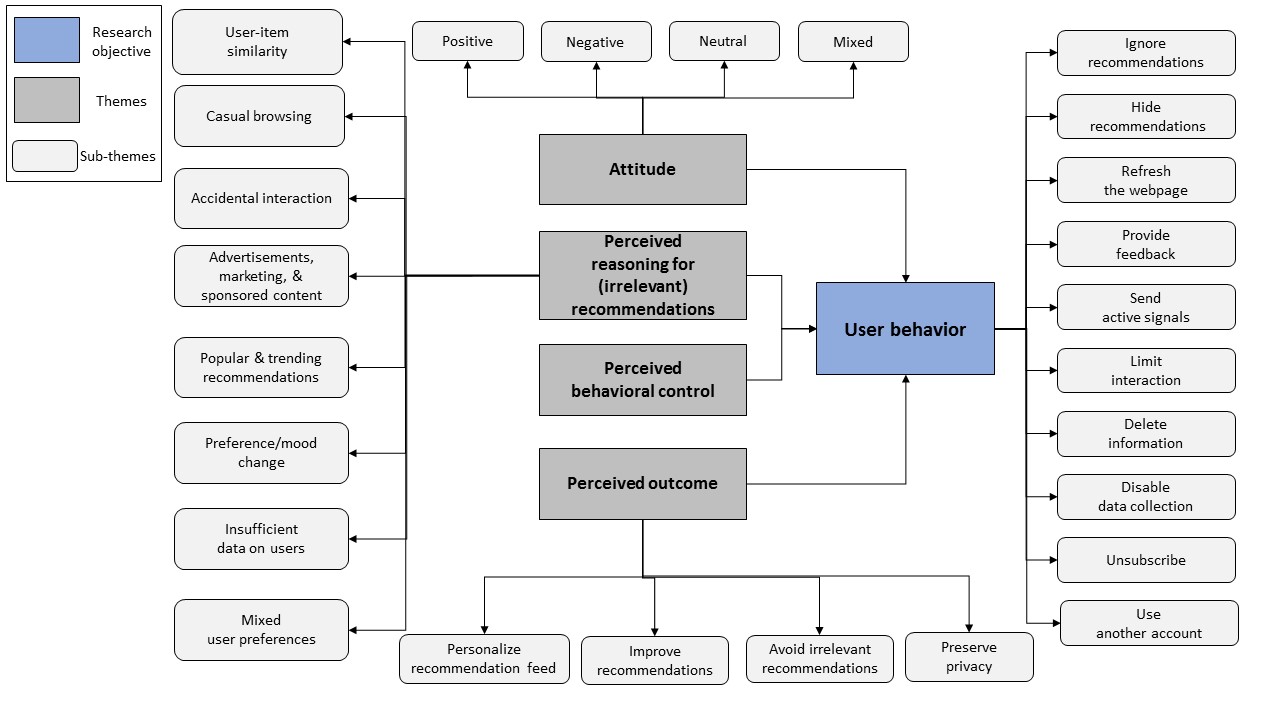}
    \caption{Thematic map summarizing the various factors that influence user behavior.} %of recommender systems. }
    \label{fig:my_label}
\end{figure}

The results of the thematic analysis are summarized in the form of themes along with its sub-themes and individual codes. 
Our thematic map present a graphical summary of the various factors that influence user behavior of the recommender systems (Fig ~\ref{fig:my_label}).
% Using the conceptual framework of the theory of planned behavior as its basis, we identified five main themes namely \textit{user's attitude towards the recommender system}, \textit{perceived reasoning for recommendations}, \textit{user behavior}, \textit{perceived behavioral control}, and \textit{perceived outcome}.
Five main themes emerged from the data analysis namely \textit{user's attitude towards the recommender system}, \textit{perceived reasoning for recommendations}, \textit{user behavior}, \textit{perceived behavioral control}, and \textit{perceived outcome}.
In the remainder of this section, we describe these themes in detail along with direct quotes and statements from the participants.
While themes are presented as discrete, some overlap of content exists between them.
% -----------------------------------------------------------------------
% -----------------------------------------------------------------------

\subsection{Attitude towards Recommender System}

% ***(edit)***

%Recommender systems actively notify users of relevant content in the form of recommendations.
%However, the inherent motivation behind these recommendations may differ from the viewpoint of consumers and providers~\citep{jannach2016recommendations}. 
%While providing users a satisfying experience is integral to these systems, from the provider's viewpoint, the recommender system's goals may be to steer user behavior in the desired direction, increase revenue, learn consumer habits, and maintain customer loyalty~\citep{jannach2016recommendations}.
%As a result, users may have varied impressions of the system depending on the nature of the recommendations they receive, and the inherent motivation behind them. 

Attitude refers to the user's overall feeling towards a recommender system~\citep{pu2011user}. 
In general, attitudes reflect the user's thoughts and opinions that determine their choices and behavior.
User attitudes can vary based on several factors.
Based on our analysis, we found that users' attitudes varied depending on the application, the nature of the recommendations they receive, perceived utility, and overall experience of interacting with the system.
In general, participants expressed dissatisfaction towards various types of recommendations that did not cater to their current needs or interests.
These recommendations, described in ~\ref{sec:Reasoning}, include targeted advertisements (personalized and non-personalized), popular or trending recommendations, and irrelevant or partially relevant recommendations.
Here, we divide this theme into four sub-themes, namely positive, negative, neutral, and mixed.
We describe these sub-themes in detail.

\subsubsection{Positive}
In general, users described the recommender system using the terms `useful', `helpful', and `convenient'. 
Participants with a positive attitude towards the system described the various benefit of using recommender systems.
They stated that the recommender system helps them find items they are looking for, discover new items, and offer a variety of options while saving them time and effort. 
One user described, 
\qte{ I think it's pretty good, [...] it's very useful for a lot of people because they don't have to always go searching for something. There's something there, recommended to them. Even if they're a new user there's things that they could look at and click on, and then it'll show them more related to that, just to get them started on the system} [P6].
Another user expressed,
\qte{.... It helps me find better deals. It helps me find out about new products and it helps me shop conveniently. So, I think it's doing a great job}[P10].

%Similarly, another user described, 
%\qte{Sometimes just getting to see other new content you know, like seeing[...] Oh I didn't know this you know ...that this was rated five stars. I have never even heard about this film or this show or a pair shoes or whatever. So yeah, I think it's of great value} [P14].

\subsubsection{Negative}
While most users conveyed a positive experience with recommender systems, a few users had a negative impression of the system.
They expressed frustration with regards to targeted advertisements, sponsored content, and their invasion of privacy. One user mentioned, 
\qte{I think [...] that it's not because they care about the customer and want to [provide]
a more personalized experience, the thing is it's just for their benefit, so they don't
really take that into account. 
I mean the [ads] are all over your feed like it's just too much. It's excessive, it's obnoxious} [P19]. % Katie Thrash,
Another user stated, 
\qte{I think it's very not private. I guess I am aware of what I signed up for. 
The app itself, I just know that all of our information is going to advertisers and it's being sold off because even on YouTube, I notice when I get advertisements sometimes it's catered to me as well} [P26].
%And I think it just depends on the person like how comfortable they are with giving their data over} [P26]. %Michelle Vu, 

\subsubsection{Neutral}
A few users held a neutral stance on the topic. 
These users were aware but indifferent to the advantages and drawbacks of using a recommender system.
We found that users with a neutral stance typically spent less time on average using recommender systems than the others.
One user expressed,
\qte{I guess I'm neutral, when it's helpful, it's nice but a lot of the times I already know what I'm looking for when I use something like Amazon or [Netflix].
So, I'm kind of just looking for what I need, I don't dislike it necessarily.
And I know that I'm being tracked in ways I'm probably not aware of, but I guess, my general thought is that like millions of people use Netflix, so it can't be that bad} [P3]. %Ashley Voss,

\subsubsection{Mixed}
A subset of users held both positive and negative impressions of the recommender system.
We also noticed that some users were uncertain or indecisive on their stance towards the system.
A user described, 
\qte{... It's dependent on the service because something like Netflix, [...] where they get their money from paid subscriptions then the recommendation service will be more geared towards trying to make sure you enjoy the product as much as possible. If it's a service that gets its money based off of ad revenue then ... it can be a negative to the user because things that grab your attention aren't necessarily good for you} [P12].

% Include a general paragraph on how user attitudes may influence behavior?

%---------- ---------- ---------- ---------- ---------- ---------- ---------- ----------
\subsection{Perceived Reasoning for Recommendations}
\label{sec:Reasoning}

The data analyzed within this theme represents the user's perception of the recommender system's knowledge, reasoning, and motivation. 
Participants described the recommendations they received from the various applications using the terms `useful', `interesting', `relevant', `somewhat relevant', `irrelevant', and `annoying'. 
Specifically, we asked participants to explain the system's reasoning for the recommendations they regarded as `irrelevant' or `annoying'.
Here, we describe their responses in detail.

\subsubsection{User-Item Similarity}
The majority of the users described that the items recommended to them were similar to what they have watched, browsed, or purchased in the past.
They further explained that the system uses this data to pick item features in order to recommend new items. 
However, these new recommendations did not necessarily align with their taste and preference.

One user described, 
\qte{It is probably based upon [...] a certain character or a certain thing that I watched and they just pulled from there and assumed that I wanted to watch another certain type of a children's movie} [P10]. %Ericka
% Use another quote???
Another Facebook user described, 
\qte{May be [Facebook] looks at certain keywords in the posts and comments that I am making. 
% I do not know. 
I tried looking through my statuses to see if there was anything that I can identify as to why Facebook thinks I am depressed 
%I just do not know why
} [P23]. %Marisa

% One user explained that the system would recommend music based on the `time period' of the songs he has listened to,
% \qte{..[The system] will sometimes recommend movies or music that I don't like.  Like, If I listened to a lot of music from the 1980s for example. It will recommend me music just because it is from the 1980s, It may not recommend music that I actually like, like specific rock music from the 1980s or specific R\&B music from the 1980s } [Male, 37, Business Owner].

% paragraph about user-user similarity. 
On the other hand, some participants explained that the recommendations were based on the preferences of other `similar' users.
While some users perceived this idea solely based on the explanations of recommendations ( for e.g., ``users who bought ... also bought ...''), other users described in detail that the system groups users based on similar interests and demographics to recommend items to the target user.
One user expressed,
\qte{Sometimes when I get those kinds of [recommendations], I would see a ... little note underneath that said “this was watched by people who watch this video”. 
But that wouldn't necessarily mean that was something that I would also want to watch} [P1]. %Alake, 
A Netflix user explained, %citing Netflix as an example,
\qte{Maybe people with similar demographics as me are interested in them, like based on what other users of similar demographics like click on and watch and like spend time on. 
The recommendation system will pull that information and suggest that} [P37].%Stephanie Hu,
\looseness=-1

Similarly, another user described in detail,
\qte{I think it probably has to do with other people in the population that fall into the similar
category as you, kind of get lumped together and they might have purchased the same
thing or something similar and they purchase another item to go along with it or looked
at another item. So, they make that suggestion based off of that or like I said people just previously making the purchase. 
So, these three things together
%So, you must want it too, kind of thing. And I may be using it for a completely different purpose, type of thing
} [P34]. %Sherri Buckner
\looseness=-1

While many participants expressed uncertainty regarding the specifics of how the system generates recommendations, these responses show that users possess an innate and general understanding of the different recommendation algorithms including collaborative filtering, content-based filtering, and model-based methods~\citep{koren2009matrix, sarwar2000application,ricci2011introduction}.

\subsubsection{Casual Browsing}
Users mentioned the idea of casual browsing, stating that sometimes they would mindlessly browse social media applications and retail websites when they feel bored. 
However, they realized that the system would track their activity such as the items they looked at, searched for, or clicked on, and the time spent browsing different items and use that information to recommend new items that they may not necessarily be interested in.
The user explains, 
\qte{... [On Amazon] I scroll through and look at the `deals of the day' which have just about everything in there. 
So, I feel like the amount of time that I spend looking at products, ...  I might click on it just to see what it is but it's not really something that I want I'm just kind of interested to see what it is used for.
And I think they store that information and then use that to make those recommendations that are not necessarily applicable to me} [P34]. %Sherri Buckner

\subsubsection{Accidental Interactions}
Several participants mentioned that the recommendations were based on items they have accidentally clicked on, or searched for in the past.
Similarly, a few users reported receiving related recommendations right after watching a certain video once or buying a particular item one-time. 
Lastly, users also mentioned that the system would recommend them similar items even after having purchased that particular product. One user described, 
\qte{I may have accidentally clicked on a product that I didn't like. Therefore, that might have altered my interests according to the recommendation system} [P32]. %Sam Jackson
Another user said,
\qte{I think it may have happened because I did click on one similar genre and then it just kept showing me that genre or something} [P7]. %Daniel
\looseness=-1

\subsubsection{Advertisements, Marketing and Sponsored content}
Participants in our study demonstrated an understanding of the goals and motivations of the recommender system.
They realize the motives behind the recommendations.
They recognize when the system is promoting a new song or pushing a certain product. 
A user described,
\qte{With Facebook, people who are paying to get their stuff out in front of more people. 
Their stuff pops up first. 
Even if it's not necessarily something I would actually like. 
They are paying for the ads} [P38]. % subject details?
\looseness=-1

Another user mentioned the notion of `marketing' stating, 
\qte{The [recommendations] may not be relevant, but they still serve it up because, this is one of their marketing functions. 
They are trying to cross promote and get you to purchase more or engage you in purchasing more. 
Or even putting that thought into your head that, “Hey, did you think about buying a bag? Did you think about buying a mouse?} [P39]. %Wayne
Similarly, one user suggested that the system is promoting certain items so they get recommended to all users, 
\qte{[...] to boost some products and services and perhaps they recommend it to everyone. 
It's ... not just specifically to me 
%but it's [recommended] to anybody who is willing to purchase or view [them]
} [P4]. %Bella, 

\subsubsection{Popular and Trending Recommendations}
Users may often find items that are popular amongst all the other users across the platform irrelevant.
These items may be products with the most sales, movies with the highest ratings, or videos with the most views.
Similar to popular recommendations, trending items focuses on popularity within a particular time frame. 
For instance, a new product released that suddenly surges in sales.
One participant shared a similar concern stating,
\qte{I think maybe ... that content is still popular amongst other users that are not like
myself maybe like overall across all the users on an application there.
So, they try to push me to try to watch it and see if I like it too. 
Even if [it is not related to my usual] habits} [P26]. %Michelle vu
Another user reasoned, 
\qte{I think some of the videos that are recommended were very popular or trending videos
that a lot of people had seen and were on my feed} [P1]. %Alake, 
\looseness-1

\subsubsection{Change in Preferences or Mood Change}
% \begin{enumerate}[noitemsep]
% \item Liked previously, no longer interested
% \item Not in the mood currently
% \end{enumerate}

Users’ interests in items may change over time.
Similarly, users may have different preferences for items depending on their current situation.
Participants conveyed a similar experience with recommendations.
One user stated, 
\qte{I was previously interested in them [items] but now I'm not interested [...] anymore .... the recommendation system may still has that type of item as I'm interested in them} [P37]. 
Another user expressed, 
% \begin{displayquote}
% I was previously interested in them but now I'm not interested in them anymore
% whether the recommendation system still has that type of item as I'm interested in
% them.
% -Stephanie Hu, Female, 19, Student
% \end{displayquote}
\qte{Yeah, YouTube would recommend me some clips of ... `American Bad' but I'm not in the mood so even like I might have been a couple days ago} [P28]. %-Patrick, 
\looseness=-1

\subsubsection{Insufficient Data on Users} 

% \begin{enumerate}[noitemsep]
% \item New user – Recommender system does not have enough information on user preferences
% \item Gauging user preferences – System is trying to see what the user likes
% \end{enumerate}
A critical challenge for any recommender system is recommending items to new users or users with an insufficient amount of activity on the platform~\citep{schein2002methods}. 
In this case, the system lacks the valuable history of the user's interaction with the system on which to base the recommendations.
Participants in our study reflected a similar understanding of this concept.
Users mentioned that the system probably does not have sufficient interaction data to recommend items.
Therefore, it is trying to understand user preferences by recommending different genres.
On the other hand, some users described the system as `inferior', stating that the recommender system may not be as good. 
Participants used the term `sliding scale' to describe the accuracy of recommendations, explaining that the recommendations may not be perfectly relevant.
One user expressed, 
\qte{I think that [...] the reason is maybe the system isn't as good, however, that looks at capturing what you are interested in, based on interaction on the platform. 
But also, I think it takes time to learn someone, and to learn someone's behavior. 
I think the more you interact on a platform, the better the recommendations, the more accurate they will be. 
Therefore, if you are getting a lot of inaccurate stuff, maybe you just haven't used the
platform that much
%Maybe you only watch two videos and so they will have a small sample size to go off of
} [P31]. %Sam clark, 
% *(edit)* Another user stated,
%\qte{I think the system is throwing out, using algorithms, testing you, to see if you would like a certain genre. 
%If you don't, it would hopefully learn to not give you the same recommendations} [P15]. %Janna, 
Similarly, another reasoned,
\qte{Maybe ... it's trying to figure out what I liked. 
It's trying to understand the user trying to understand what type of products or what type of music or whatever, like user understanding} [P13]. %Gabrielle, 
\looseness=-1

\subsubsection{Mixed user preferences}
A user who shares his user account with his family members reasoned that he finds the recommendations irrelevant, stating
\qte{[...] I share a Netflix account with another family member and there's recommendations because you know they are on the same account.
You know using, searching and watching. 
So, it's kind of their preferences mixed in the recommendations} [P7].  %Daniel

%---------- ---------- ---------- ---------- ---------- ---------- ---------- ----------
\subsection{User Behavior}
\label{sec:Response}

The data analyzed within this theme represents how different users interact with recommender systems.
Our analysis suggests that users interact with the system differently based on several factors.
These factors include the user's behavioral intentions, their knowledge of the recommenders, and their attitude towards the system.
We found that users with strong attitudes (positive, negative, and mixed) develop behavioral intentions.
These behavioral intentions together with the user's knowledge of the available system features, their perceived behavioral control, and their beliefs about action-interaction outcomes guide their actual behavior.
In this section, we present the different ways users respond to recommendations along with their intentions.
\looseness=-1

\subsubsection{Ignore recommendations or Hide recommendations}
The majority of the users mentioned that they `do nothing'. %  when they receive irrelevant recommendations. 
A user described, 
\qte{I usually just ignore them or like, I use the app and I get like notifications and I know I can turn it off, but I am kind of lazy about stuff like that, so I just kind of discard it or dismiss it or like scroll past it} [P3].  %Ashley Voss
On the other hand, few users mentioned that they view their recommendations stating that the recommendations have helped them discover something new and interesting in the past.
A user expressed,
\qte{I look at it. Just see what it's all about and sometimes this is a good way to discover 
something I have never used or seen before also} [P4]. %Bella Volksy

Some participants reported using the application features to explicitly communicate their dislikes.  
One user mentioned, ``I will hit `hide ad' and `don't show notifications' of this ad or something like that.''
Our analysis shows that users are aware of the various applications' feedback mechanisms, and use them to signal their preferences to the system. 
% \begin{enumerate}[noitemsep]
% \item Don’t show me this again
% \item Close a category (YouTube)
% \end{enumerate}
One user mentioned, 
\qte{I usually ignore them or If I don't want to see it in my
recommended section, [On YouTube] I could click and say I'm not interested in this [...]then they usually take the video away} [P9]. % Destiny
A Facebook user expressed,
\qte{I click on the x icon 
%on the upper right-hand corner, 
[...] and then afterwards it [asks] 
%shows a bar that says, 
%``do you want to stop receiving information from this again?'' And then I click yes and it asks 
why? I click irrelevant and then it says you will no longer receive recommendations like this in the future} [P17].

%\qte{I click on the x icon on the upper right-hand corner, [...] And then afterwards it shows a bar that says, ``do you want to stop receiving information from this again?'' And then I click yes and it asks why? ... I click irrelevant and then it says you will no longer receive recommendations like this in the future} [P17].

%\subsubsection{Hide recommendations}
%Some participants reported using the application features to explicitly communicate their dislikes.  
%One user mentioned, ``I will hit `hide ad' and `don't show notifications' of this ad or something like that.''
%Our analysis shows that users are aware of the various applications' feedback mechanisms, and use them to signal their preferences to the system. 
% \begin{enumerate}[noitemsep]
% \item Don’t show me this again
% \item Close a category (YouTube)
% \end{enumerate}
%One user mentioned, 
%\qte{I usually ignore them or ... If I don't want to see it in my
%recommended section, [On YouTube] I could click and say I'm not interested in this [...]then they usually take the video away} [P9]. % Destiny
%Another Facebook user expressed,
%\qte{I click on the x icon on the upper right-hand corner, [...] And then afterwards it shows a bar that says, ``do you want to stop receiving information from this again?'' And then I click yes and it asks why? ... I click irrelevant and then it says you will no longer receive recommendations like this in the future} [P17]. %Joshua

\subsubsection{Refresh the page for new recommendations}
A few participants mentioned refreshing the page until they see recommendations they would like.
One user mentioned, \qte{I usually just like refresh the page which always gives me new  recommendations and I will just refresh it until I find something I like}[P28].
Another user described,
\qte{... [On YouTube], I usually just ignore it or like refresh to have ... a new set of recommended videos. 
For Amazon, I will just keep scrolling and move on.  
Sometimes if I don't like what they are recommending, I just filter more to have like those type of things} [P34]. % Hamza, 

\subsubsection{Provide feedback}
% Quote with the fill out surveys?
Data analyzed within this sub-theme shows that users communicate their preferences to the recommender system through various interactions. 
These interactions include liking or disliking a video, rating items, filling out surveys, creating a music playlist, etc. 
Users believe that these interactions allow the recommender system to recognize their actual preferences and improve the recommendations. 
On the contrary, we also found that some users habitually upvote, downvote, or rate items to express their disposition without any behavioral motivation. 
A video streaming service user expressed, 
\qte{I actually do [press like or dislike] a lot on YouTube because it helps to really predict what's going to play and not play. So, I hate when things play that I don't like} [P18].
\looseness=-1
A user described, 
\qte{On Google play, If I'm listening to a radio that's very specific .. and I ... ask for workout music. 
It'll play the music but as soon as I click a thumbs down it'll stop the song, skips the songs and then like never play the song again. 
I click the thumbs down button or if there is anything that says like I don't want to see this anymore then I'll go ahead and do that} [P14].
In contrast, one user expressed,
\qte{I usually just add them to ... my favorite playlist. I like [press like] them %I don't comment that often. I usually just read the comments...
}. 
When asked if she had any motivation for those interactions, she further explained,
\qte{
I never really thought about that, I think it probably does but I just always assumed as long as I click on a video to watch it then they would just like show me another video that's related} [P26].

% \begin{enumerate}[noitemsep]
% \item Press like
% \item Press dislike
% \item Rate items
% \item Fill out surveys
% \end{enumerate}

% Need another quote.

\subsubsection{Send active signals}
\label{subsec:activesignals}
A subset of users explained that sending active signals to the recommender system helped them avoid irrelevant recommendations.
These signals include `actively' searching for items, clicking on items, adding or removing items from a playlist, or changing one's preference profile to signal the recommender system.
This is different from the above sub-theme `providing feedback'.
Here, the user is intentionally spending a session sending interactive signals.  
They expressed that these actions enable the recommender system to realize their actual preferences and help steer the recommendations in the desired direction.
One user stated, 
\qte{Probably change the profile, ... like let them know [what kind of things you like]}[P30].
Another user mentioned sending signals to the recommender system, 
\qte{I try to skip through the [recommendations], or I will actually actively pick something that is the direct opposite of it. 
So, [the system] would be like, oh okay you like MSNBC not Fox News
%, I'm sorry. Right, I try to send it a direct obvious signal
} [P29]. %Rachel, 
Some users mentioned adding items to their playlist on a music streaming service to signal the system about their preferences.
For instance, a user expressed,
\qte{I [press] like, dislike or `favorite' [on Spotify], or I can interact in a different way [...]
%that I want to say like for example, adding something to my own playlist or sharing something too. 
I think it keeps track of that} [P22].%Michelle

\subsubsection{Limit interactions}
\vspace{-1mm}
% Avoid searching - (London, Srikant, shadana )

A subset of users mentioned that they refrain from casually interacting with the recommender in a way that would incorrectly signal the system about their interests. 
Furthermore, they described that the system tracks their activity on the platform.
Therefore, they refrain from interacting with any recommendations unless they align with their interests.
A user expressed, 
\qte{I think a lot of your user interaction is tracked through the number of clicks,  the number of searches you performed, types of searches you performed, ... the departments you have looked into. 
If [a user] does not like the recommendations, I guess they should just avoid searching for that product. 
... I mean it's kind of in the user's control} [P25]. % Srikant, 
Similarly, another user described,
\qte{... If I search something new something out of what I revolve around maybe. %let’s say, a saddle. 
I don't have a cat, I certainly am searching cat food for a friend, ... I feel like searching things kind of effects it, so I guess avoiding it} [P22]. %London, 
\looseness=-1

\subsubsection{Delete information}
\vspace{-1mm}
% \begin{enumerate}[noitemsep]
% \item Browsing history / Browser history
% \item Search history
% \item Cookies
% \item Remove items from playlist 
% \end{enumerate}

Participants reported deleting items from their browsing history, search history, and playlists.
Similarly, some users mentioned that they remove items from their `watch later' list. 
When asked about the reason for these actions, users responded that the recommender system likely uses this information to generate new recommendations. 
One user described, 
\qte{I go to my history and delete my search list and then I unsubscribe to some of the
channels and then I remove a lot of videos from my watch later. 
That helps me like you know like by 50 \% may be} [P13]. % Shadana, 

%Another user expressed, 
%\qte{
%I don't know if cleaning out your search history on your computer does anything, I do that fairly often. 
%I know [on Facebook] that there's probably ways to change your privacy in which information is stored on you and so on. 
%I do a little bit with that, but I don't do that very often but I kind of clean out my search history and things that they have saved that I have looked at} [P34]. % Sherri Buckner,  
\looseness=-1

\subsubsection{Turn off data collection}
A few users mentioned the idea of turning off their browsing history or watch history. 
While some users were certain about the outcome of these behaviors, other users vaguely expressed this idea.
One user explained,
\qte{For something like ... internet websites, ... I go through stuff like my Google account settings to turn off all the personalized ad revenue and data collection because I personally do not enjoy having lots of data collected about me by faceless organizations} [P12]. %Ford
Another user discussed, 
\qte{I think you can turn off notifications. 
I don't know, I know on Amazon you can turn off the browsing history and then, I'm guessing [...] even if you don't see a button, you may be can like write to them.
Um, I usually just ... ignore it versus trying to turn it off
%, so I haven't looked into it
} [P3]. %Ashley

\subsubsection{Unsubscribe}
A few users of streaming services like YouTube and Spotify mentioned the idea of unsubscribing from artists and channels.  
They described that unsubscribing from certain channels or artists helped them avoid annoying and irrelevant recommendations.
For instance P15 stated, 
\qte{On YouTube, I unsubscribe to the channel if I'm getting really annoying notifications about it}.
Another user stated, 
\qte{... [Deleting search history] doesn't help really. It helps me when I unsubscribe to the channels and then delete it from my watch later} [P13]. % Shadana, 
%A third stated,
%\qte{ ... At one point, I kept getting an advertisement for a channel that I had never watched before. I watched it [video] and I didn't like it so I disliked it. I still kept getting recommended the channel.[..] you could actually click on three dots next to the recommended video and say, “Hey, I'm not interested in this”. I would do that, and it would still recommend that channel to me. Eventually I just ended up blocking the user and the channel. Eventually it went away. So yes, I'm sure there's plenty of flaws in the system on that front} [P5].

\subsubsection{Use another account}
\vspace{-1mm}
A couple of users also mentioned using a different account to avoid irrelevant recommendations.
One user explained,
%that leaving the account passive for a while helps getting rid of certain recommendations.
\qte{... I have 2 accounts, so, I just try using the other one for a while and leaving the other one passive} [P15]. %Shadana, 

%---------- ---------- ---------- ---------- ---------- ---------- ---------- ----------
\subsection{Perceived Behavioral Control}
\label{sec:control}
\vspace{-1mm}
% What is perceived behavioral control?

Perceived behavioral control refers to the user's perception of the ease or difficulty of performing the behavior of interest.
%According to Bandura~\citep{bandura2010self}, ``it is concerned with judgments of how well one can execute courses of action required to deal with prospective situations.''
According to the theory of planned behavior, perceived behavioral control, together with behavioral intention, can be used directly to predict behavioral achievement.
However, the likelihood of a given behavior is also governed by, to some extent, the resources and opportunities available to a user.

Our analysis of the participants' responses shows that users, based on their understanding and interactions with the system, have ascertained several resources that enable them to indulge in the aforementioned behaviors.
These resources include the use of available system features to provide feedback to the recommender system such as rating items, liking or disliking a video, filling out surveys, etc.
However, users have also discovered unconventional methods to obtain desired recommender system behavior.
These methods include turning data collection off, deleting user activity data like browsing history, search history, or purchase history, actively interacting with items of interests, and limiting interactions. %  such as liking or disliking a certain genre of movies and music
Users' perceptions of behavioral control varied across the sample. 
Most notably, we found that users' engagement in behavior was largely impacted by their intentions.
Some users, possessing the knowledge of available resources and action-interaction outcomes, chose not to carry out these behaviors solely based on their intentions, while other users demonstrated carrying out these behaviors in a variety of situations.
\looseness=-1

%---------- ---------- ---------- ---------- ---------- ---------- ---------- ----------
\subsection{Perceived Outcome}
\vspace{-1mm}
\label{sec:Outcome}
Perceived outcome refers to the users' beliefs about the expected outcome of their interactions with the system~\citep{ajzen1991theory}.
Such beliefs are formed based on the users' experience of interacting with the system and drawing inferences over time.
Our analysis shows that users held various perceptions of the outcomes of their behavior. 
Based on the anecdotal evidence presented in ~\ref{sec:Response}, it can reasonably be inferred that some users attempt to steer the recommender system's behavior in desired directions.
 % - Personalize their feed themselves.  
As such, we observed most participants in our study believed that providing feedback through various interactions such as liking or disliking videos, rating products, adding or deleting songs from a playlist helps improve the quality and accuracy of their recommendations.
Conversely, some users perceived that limiting interactions with the system, and aggressively upvoting or downvoting items allows them to avoid irrelevant recommendations. %  from the system.
Similarly, some users asserted that they could personalize the recommendations to their liking by controlling the information they allow the recommender system to perceive. 
Others reasoned that these behaviors enabled them to achieve a diverse and interesting recommendation feed.
% - Preserve privacy - ford
Some users held a positive outlook towards the system tracking their activity.
They believe that the data is used to run the applications and improve recommendations.
On the contrary, other users believed that deleting user activity data on the application, turning off data collection features, or using another account would preserve their privacy.
% Users think the system is not good enough.
%---------- ---------- ---------- ---------- ---------- ---------- ---------- ----------

\section{Limitations}
\vspace{-1mm}
% Structure
Our study has two noticeable limitations. 
%Despite various measures to ensure the quality of our research, our study has two noticeable limitations. 
First, subjects were recruited from Reddit. 
These individuals are likely more technologically savvy than the average internet user.
Second, interview data were analyzed by coders with a background in recommender systems.
However, to overcome researcher bias and ensure reliability, two researchers analyzed the data independently.
Any coding discrepancies between the researchers were reconciled by consensus decisions.
Despite these limitations, this study provides strong subjective evidence of the presence of unconventional user behavior across many domains of recommender systems.
Finally, we discuss future work to further explore this research direction.

%----------------------------------------------------------------

\section{Discussion}
\label{sec:Discussion}
\vspace{-1mm}
%----------Structure--------------

% Summary of the whole paper in 3 sentences.
Our findings, based on a rigorous thematic analysis, demonstrate that users possess an intuitive and sophisticated understanding of the recommender system's behavior.
We observed that user attitudes were influenced by the user's perception of the utility, and overall experience of interaction with the system including the intrinsic nature of the recommendations served to the user.
The study showed that the user's conceptualization of how recommender systems work has a significant influence on their behavior.
In summary, we found that user behavior is informed by the user's beliefs and knowledge of the action-interaction outcomes, perceived behavioral control, their intentions and attitude towards the system.
% During our analysis of the collected data, we observed themes of end-user debugging and control. 

%\subsection{How does end-user debugging and control impact recommendation performance?}
%Recommender systems rely significantly on user interaction data to accurately predict user preference.
%Therefore, an incorrect interpretation of user preference may result in user dissatisfaction.
%The concept of end-user debugging~\citep{kulesza2012tell, kulesza2010explanatory} involves the user mindfully and purposely adjusting the system's reasoning so it aligns with their actual preferences.

%In this study, we observed many participants express dissatisfaction towards various types of recommendations that did not cater to their current needs or interests.
%Participants described the system's reasoning in detail (section ~\ref{sec:Reasoning}), explaining how the system may have incorrectly interpreted their preference for items.
%In summary, their responses mentioned that the recommendations were based on casual browsing sessions, accidental interactions, mixed user preferences, or simply their preference has changed.
%Others reasoned that the recommendations based on popularity, trends, user-item similarity, and targeted advertisements did not accurately capture their preferences. 
%Their responses indicate that users possess an intuitive understanding of the recommender system's behavior.

We analyzed user interaction with recommender systems, specifically how they respond to irrelevant recommendations (section ~\ref{sec:Response}).
%Participants expressed a latent concept of control over their recommendations~\citep{tintarev2011designing}.
Throughout the interviews, we observed three contrasting approaches of end-user debugging and control used by the participants~\citep{tintarev2011designing, kulesza2012tell, kulesza2010explanatory}.
In the first case, participants were not fully aware of all the resources available to them to provide feedback to the system.
For instance, P26 did not realize the function of a like button, however, she described using the `not interested' feature to express her preference.
P26: \qte{I never really thought about that ['like' function on YouTube]. 
%I think it probably does but 
I just always assumed as long as I click on a video to watch it then they would just like show me another video that's related. [...] Yeah I think like sometimes YouTube recommends me scary videos that I just don’t want to see it. There is usually an option for me to take it off so I [...] click that}[P26].
Therefore, users who does not possess the knowledge of the system's functionality may not be able to enjoy the intended utility of recommenders.
\looseness=-1

Similarly, some participants expressed a mature understanding of how feedback functions work and used them accurately to signal their preferences, P29 explained in detail,
\qte{On Pandora [...] basically by saying thumbs up, you're saying I want to hear more like this. Now, it’s not necessarily by the same artist, they'll actually take a look at the components of the song. Like, does it have a Caribbean beat, is it a rap song or a musical? So, you have to be careful because sometimes if a band is doing a cover of something then you're like oh I don't really like that cover they just did, you thumbs down it %because you just don't wanna hear that cover. 
And then all of a sudden that band disappears from your playlist.
%you're like No..... 
So, apparently my wife informs me you can actually pick I don't want to hear this track versus thumbs downing that, but it is not intuitive and difficult to find.}

In contrast, few participants described using unconventional interaction methods to avoid irrelevant recommendations, as described by P13, 
\qte{[On YouTube] I've tried clearing my cookies, that didn't work. Well I would just not search for it for a while. I go to my history and delete my search list and then I unsubscribe to some of the channels and then I remove a lot of videos from my watch later. That helps me like you know like by 50 \%} [P13].
We speculate that such user behavior may have an adverse affect on recommendation performance and indirectly affects the user's perceived usefulness, trust, and overall satisfaction with the system.
% *(edit)* Our study provides subjective evidence that user behavior is largely influenced by their knowledge of the system, behavioral intentions, and their attitude toward the system.

% \subsection{Implications}

Our research has several implications. % in both theory and practice.
%This research has several implications in both theory and practice.
First, we encourage researchers to establish the link between the user's mental model of the recommender system and their online behavior. 
%Understanding the user's mental model of how recommender systems work will help obtain a refined understanding of user behavior and identify unconventional behaviors such a mental model would elicit. 
Since user behavior is both input for recommendation algorithms and constrained by them, unconventional user behavior may potentially be detrimental to recommendation performance.
Second, designing system that can identify and leverage these behavior can significantly improve recommendation performance.
Third, we suggest examining discrepancies between the users’ mental model and the system’s actual behavior to identify incorrect user beliefs and assumptions about the system's functionality. 
Therefore, resolving the identified discrepancies can help inspire trust, confidence, and satisfaction with the system.
Finally, we suggest the use of explanations to help users fully comprehend the functionality of feedback functions and implement controllable interfaces to allow users to revise their preferences in an accessible and intuitive manner.
\looseness=-1

\section{Conclusion and Future Work}
\vspace{-1mm}
\label{sec:Conclusion}

% In this paper, we explored the users' attitude towards the recommender system, their perception and reasoning of how the system operates, their online behavior, and the motivation behind them.
In this paper, we identified and explored the various factors affecting the user behavior of the recommender systems.
To that end, we interviewed forty recommender system users from Reddit in a qualitative user study.
Our analysis of their responses demonstrates that user's behavior towards the system can be influenced by their attitude towards the system, their perception and reasoning of how the system operates, and their motivation.
Our findings contribute to a refined understanding of user behavior and demonstrate the relationships among the several factors affecting it. 
For future work, we plan to develop a comprehensive theoretical framework of the users' mental models of the system.
We imagine a framework that will help recommender systems to identify and leverage these behaviors to enhance the users' experience and improve the system's performance.
Finally, we plan to experimentally evaluate the impact of the identified user behaviors in a variety of domains.

\bibliography{main}

\end{document}